\documentstyle[12pt]{article}

\begin{document}

\begin{flushright}
 LYCEN 9713
\end{flushright}

\begin{center}

\vskip 2.cm

{\bf \large WHAT DO EXPERIMENTAL DATA "SAY" ABOUT GROWTH
OF HADRONIC TOTAL CROSS-SECTIONS~?}

\vskip 1.5cm

{\bf P. Desgrolard $^{1}$, M. Giffon $^{1}$, E. Martynov $^{2}$       }\\
\vskip 0.5cm

$^{1}$ Institut de Physique Nucl\'eaire de Lyon, \\
IN2P3-CNRS et Universit\'e Claude Bernard, \\
F 69622 Villeurbanne Cedex, France\\
\vskip 0.3cm
$^{2}$ Bogolyubov Institute for Theoretical Physics,\\
National Academy of Sciences of Ukraine\\
Metrologichna st., 14b, 252143 Kiev, Ukraine\\
\vskip 2.cm

\end{center}

{\bf Abstract} We reanalyse $\bar p p$ and $pp$ high energy data of the
elastic scattering above $\sqrt{s}=5$ GeV
on the total cross-section $\sigma_{tot}$ and on the forward $\rho$-ratio
for various models of Pomeron, utilizing two methods.
The first one is based on analytic amplitudes, the other one relies on
assumptions for $\sigma_{tot}$ and on dispersion relation for $\rho$.
We argue that it is {\bf not} possible,
from fitting only existing data for forward scattering, to select a definite
asymptotic growth with the energy of $\sigma_{tot}$.
We find equivalent fits to the data together with a logarithmic Pomeron
giving a  behavior $\sigma_{tot} \propto \ln ^\gamma s$, $\gamma\in [0.5,2.20]$
and with a supercritical Pomeron
giving a  behavior $\sigma_{tot} \propto s^\epsilon $,
$\epsilon\in [0.01,0.10]$.

\newpage

{\bf 1 Introduction}
\medskip

The question "How fast do total cross-sections of hadronic interaction rise~?"
is discussed permanently (see for example \cite {Blois} and references therin).
A definite answer to this  question, correlated to
that of the prediction of the asymptotic growth is still lacking.
The Tevatron is at the lower edge of the asymptotic range, while the cosmic
rays data (because of large uncertainties) may serve only as a guide.
Interesting information on the energy dependence of the total
cross-sections are expected from future experiments projected near the
new generation of colliders,
for example at the LHC \cite{TOTEM} with 10 to 14 TeV center of mass energies,
where, presumably, the asymptotic range will have been reached.

There is a number of theoretical and phenomenological
models together with fitting procedures using various
selections of data that
lead to controversial predictions about an asymptotic growth of the
total cross-sections. To avoid an overlong list of papers devoted to this
subject, we do not give the references for all original papers. We only note
that the main prediction varies from a very slow growth when the
Mandelstam variable
$s\rightarrow\infty$ $$\sigma_{tot}\propto \ln^{\gamma}(s/s_0), \qquad
\gamma \simeq 0.4                                                   \eqno(1)$$
as for the  "critical" Pomeron in the Reggeon
Field Theory \cite{RFT}, up to a very fast growth
$$\sigma_{tot}(s)\propto
(s/s_0)^\epsilon, \qquad \epsilon \simeq 0.1                         \eqno(2)$$
as in the model of the "supercritical" Pomeron, popular with the work \cite{DL}
of Donnachie-Landshoff (DL); of course the latter "overunitary" behavior
should be changed at very high energies to respect the
Froissart-Martin bound \cite{FM}
$$\sigma_{tot}\leq \frac{\pi} {m^2_{\pi}} \ln^2(s/s_0).             \eqno(3)$$
As a rule, each hypothesis exhibits a good agreement with the
experimental data. At the same time, an accurate comparison of the quality of
the fits given by different models is not easy because often a single
model is discussed in details with a specific set of experimental data.
Only a comparison between models that uses a
common and complete set of experimental data is abble to bring an
answer to the question in the title of this note.

Two methods are intensively investigated to analyse the experimental data on
the total cross-sections ($\sigma_{tot}(s)$) and on the ratio of the real to the
imaginary part of the forward scattering amplitude ($\rho(s)$)~:

-1) the first one uses an explicit analytic form of the complex scattering
amplitude $A(s,t=0)$

-2) the other method is based on phenomenological assumptions on the  total
cross-section $\sigma_{tot}(s)$ complemented by the dispersion relation
between the real and  imaginary part of the forward elastic amplitude. Because a
nonvanishing cross-section is assumed with increasing energy, a dispersion
relation with one substraction constant is used.

In both methods, the free
parameters are determined by a fit to the experimental data.
{\it A priori} they should lead to similar (if not identical)
conclusions for the behavior of $\sigma_{tot}(s)$. Unfortunatly, it is not the
case. As an example, let us quote first Ref. \cite{DGLM}, where the
 method using analytic amplitude was applied for the
analysis of data~:
"It is found that available experimental data
at $t=0$ do not seem to indicate a growth of the total cross-sections
faster than the first power in logarithm of the energy.". Let us quote also
a recent paper \cite{VB} where the approach using dispersion relation is
investigated in the same energy range~:
"For the whole set of present data statistical analysis ($\chi^2/d.o.f.$)
seems to favour a "Froissart-like" (logarithmic $(\ln s )^{\gamma\approx 2}$)
rise of
the total cross-section rather than a "Regge-like" (power $s^{\epsilon}$) one".
A similar conclusion holds in \cite{H}.
Thus, there is a certain contradiction between the conclusions obtained
in two approaches of analysing the data.
One may find many other recent examples of contradictory conclusions~: for
example, Block and White \cite{BMW} are
supporters of $\ln s$, while Kang and Kim \cite{KK} argue that models with
$\ln s$ and $\ln^2 s$
are equally compatible with the data; Bertini {\it et al.} \cite{Bertini}
from an analytic evaluation of the real part of the forward scattering
amplitude, based on the dispersion relations, report equivalent
success for $\ln^2 s$ and $s^\epsilon$ behaviors; Cudell {\it et al.}
\cite{CKK} exhibit the success of a DL-like model....

This note is a tentative to clear up this problem. For that purpose, we
analyse the current experimental data by the above mentioned methods using
a common set of data. We include in our analysis all published data
\cite{PDG}, excluding only those at low and intermediate energies with an
error greater than 1 mb and those which were corrected
by experimentalists in a later publication. Finally, we are left with a grand
total of 187 non filtered data distributed as follows~:
for total cross-sections of $pp$ (72 points), $\bar p p$ (47
points)-scattering and for the ratio ($\rho$) of real to imaginary part of the
forward amplitude (correspondingly 55 and 13 points)
at the energies
$$5{\hbox{ GeV}} \le \sqrt s \le 1800{\hbox{ GeV}}.$$
Of course, such a set favors low energy data (especially $pp$),
but from a theoretician point of view
it seems to us hazardous to omit some experimental data. However,
to satisfy a suspicious reader, unwilling to manage
with the discrepancies at the Tevatron
between CDF and E710 results, we also discuss the consequences of
excluding the (five) values above 546 GeV from the data set.

To be complete, we explored formulations giving two alternatives
for the asymptotic behavior of $\sigma_{tot}(s)$~: logarithmic (as in (1)) and
power-like (as in (2))  rising with energy.

\bigskip
{\bf 2  "Analytic amplitude" and "dispersion relation" fits}

\medskip
-1) {\bf Analytic amplitude fit.}
\smallskip

The (Born-level) amplitude for elastic forward scattering is the
explicit analytic function of the center of mass energy $\sqrt{s}$
$$A^{\pm}(s,0) = P(s) + F(s) \pm \Omega(s), \eqno(4)$$
where $+(-)$ stands for $\bar p p (pp)$ respectively and
where, as usual in the Regge approach, the secondary ($f$- and $\omega$-)
Reggeon contributions are
$$F(s) = ig_f(-is/s_0)^{\alpha_f(0)-1},                            \eqno(5)$$
$$\Omega(s) = g_{\omega}(-is/s_0)^{\alpha_{\omega}(0)-1}.          \eqno(6)$$
For the Pomeron contribution, we consider two parametrizations.
The first one is a standard picture for a logarithmic Pomeron,
(with a unit intercept, leading to a logarithmic rising asymptotic total
cross-section)
$$P(s) = i[g_0 + g_1\ln ^{\gamma}(-is/s_0)].                        \eqno(7)$$
The second one is a generalization of the DL
model \cite{DL} for the supercritical Pomeron (with an intercept
$\alpha_p(0)=1+\epsilon$, leading to a power rising asymptotic total
cross-section)
$$P(s) = i[g_0 + g_1(-is/s_0)^{\epsilon }] ,\quad\epsilon > 0  ,   \eqno(8)$$
where the constant term we add stands for a preasymtotic contribution.
We have also investigated the original DL parametrization with identical
Reggeon intercepts $\alpha_f(0)=\alpha_{\omega}(0)$ and a simplified
Pomeron $g_0=0$. We
concluded as in \cite{DGLM,VB} that this option, leading to a too high
$\chi ^2$, is not suitable in the present energy range and should be rejected
on behalf of the general case $g_0\neq 0,\,\alpha_f(0) \neq \alpha_{\omega}(o)$.

We satisfy the optical theorem (normalization) with
$$\sigma_{tot} = \sigma^{\pm}(s) =8\pi\ \Im\hbox{m} A^{\pm}(s,0)     \eqno(9)$$
and set everywhere $s_0=1$ GeV$^2$.  \\
The calculation of the $\rho$-ratio is straigtforward from its definition
$$\rho^\pm (s)= {\Re\hbox{e} A^\pm(s,0)\over\Im\hbox{m} A^\pm(s,0)}\ .
                                                                    \eqno(10)$$

Within this formulation, we are left with 7 free parameters ( 4 "coupling"
constants, 3 intercepts --or power--) to be determined by a $\chi^2$
minimisation (from the fit to the
experimental data on $\sigma^{\pm}$ and $\rho^{\pm}$).
We note that the CERN standard computer program for $\chi^2$ minimization
"MINUIT" does not "feel" a sensitive dependence of amplitudes with respect to
$\gamma$ in (7) --and $\epsilon $ in (8)--. Actually, for all
investigated cases, we found an output value of this parameter very closed to
the input one. Therefore, we adopt a procedure which consists
in fitting the other 6 free parameters for
fixed values of $\gamma$ --or $\epsilon $--  and then plotting $\chi^2$ versus
$\gamma$ --and $\epsilon $-- to deduce the optimal value.
This is time-consuming, mainly because it is not straightforward to find the
minimum  $\chi^2$ for a given  $\gamma$ --or $\epsilon $--, but in our opinion,
it is the price to pay to extract a valuable conclusion.

\medskip
-2) {\bf Dispersion relation fit.}
\smallskip

In this method the total cross-sections $\sigma^{\pm}$ are
parametrized explicitly.
The phenomenological expressions for $\sigma^{\pm}$ are defined in
Refs.\cite{VB,H,A} as
follows  (we use the same notations for the relevant terms as in the
analytic form of amplitudes in order to emphasize the similarities
and to permit a transparent comparison of results)
$$\sigma^{\pm} = \sigma_P + \sigma_f
\pm \sigma_{\omega},                                               \eqno(11)$$
with
$$\sigma_f = g_f(E/E_0)^{\alpha_f(0)-1},                           \eqno(12)$$
$$\sigma_{\omega} = g_{\omega}(E/E_0)^{\alpha_{\omega}(0)-1}       \eqno(13)$$
and first of all
$$\sigma_P = g_0 + g_1\ln ^{\gamma}(s/s_0).                        \eqno(14)$$
In relation with the expected smallness of $\epsilon$, we also consider
the supercritical Pomeron model (generalized DL form)
$$\sigma_P = g_0 + g_1(s/s_0)^{\epsilon}.                          \eqno(15)$$
The ratios $\rho^{\pm}$ of the real to imaginary
part of amplitudes are calculated from the dispersion relation \cite{S}
$$\rho^{\pm}(s)\ \sigma^{\pm}(s)=\frac{B}{p}+\frac{E}{\pi p}\int
\limits_{m}^{\infty}dE'p'\bigg [\frac{\sigma^{\pm}(s')}{E'(E'-E)} -
\frac{\sigma^{\mp}(s')}{E'(E'+E)}\bigg ].                           \eqno(16)$$
In these equations $E(p)$ is the proton energy (3-momentum) in the laboratory
system, $E_0=1$ GeV,
$m$ is the proton mass and $B$ is a substraction constant.
Thus, the real part of the amplitudes depends on the asymptotic cross-sections.

We note that, for historical reasons of convenience, the contributions of the
secondary Reggeons and of the Pomeron include different variables (namely $E$
for the Reggeons in (12,13), $s$ for the Pomeron in (14,15) and
$E_0=1$ GeV does not correspond to $s_0=1$ GeV$^2$).
We call this mixed formulation a "ES-parametrization".
In order to perform a complete comparison of various models, in addition we
also considered more coherently that we call a "SS-parametrization" and a
"EE-parametrization". In the SS-parametrization, the variable $E/E_0$ in (12)
and (13) is changed
for $s/s_0$. In the EE-parametrization, the variable $s/s_0$ as well as
$E/E_0$ are changed for $E/m$ (it is necessary to use $m$ instead of
$E_0=1$ GeV because  $\ln
E/E_0 \le 0$ for  $m\le E \le E_0$ implies $\sigma_P$ complex for $\gamma\ne
1$ or $2$).

Within this second formulation, we have 8 free parameters (4 "cross-section"
constants, 3 intercepts --or power-- and the substraction constant) to be
determined by a $\chi^2$ minimisation.
The same remark as above for the determination of  $\gamma$ --and $\epsilon$--
holds, consequently  we adopt a similar point of view~:
minimizing over the other 7 free parameters for
fixed values of $\gamma$ --or $\epsilon$-- and then
deducing the optimal $\gamma$ --and $\epsilon $--
value from the plot $\chi^2$ versus  $\gamma$
--and $\epsilon $--.
                               \vfill\eject
\bigskip
 {\bf 3  Results and discussion.}
\medskip

A selection of our results of $\chi^2_{d.o.f}$ versus $\gamma$ and
$\epsilon$ ({\it i.e.} for a logarithmic and a supercritical Pomeron)
is displayed in Figs.1a-b for the analytical amplitude fits
and in Figs.2a-b for the dispersion relation fits.

We find (see Fig.1) that the $\chi^2_{d.o.f}$ issued from the truncated data
set (with $\sqrt{s}\in [5, 546]$ GeV) is slighly better than the
$\chi^2_{d.o.f}$ issued from the entire data set (with
$\sqrt{s}\in [5,1800]$ GeV)~: 1.10 instead of 1.15, the two sets giving
homothetic curves.
However, these differences are not significant
(the corresponding parameters yield
undistinguishable plots of the cross-sections and of the $\rho $-ratios);
they are due essentially to the data in conflict at high energies.
This trend is exhibited only in the case of analytical fits but might be
extended to dispersion relation fits.                        \\
We find also (see Fig.2) that the $\chi^2_{d.o.f}$
are quite comparable in the three (ES, SS, EE) parametrizations.  \\
Furthermore, the examination of Figs.1 and 2 does not allow to find a
signi\-ficant hierarchy
\par ({\it i}) between the logarithmic Pomeron and the supercritical Pomeron
\par ({\it ii}) and, to a lesser extend, between the first method using
analytical amplitudes and the second method using
phenomenological cross-sections and the dispersion relation;
we only note the second one gives a minimal
$\chi^2_{dof}\simeq 1.10-1.11$, slightly better than
$\chi^2_{dof}\simeq 1.14-1.15$ for the first one.
Let us mention also that this second method would give as an indication
$$\gamma\sim 2.2 \hbox{   and  } \epsilon\sim 0.07\ .$$

However, all of these figures  (in spite of a large vertical scale) display a
wide plateau in the $\chi^2$ behavior.
So, the proposed values must not be separated from the wideness of this plateau.
The first method is somewhat more selective to exclude highest values
of $\gamma $ and $\epsilon $,
while the second one excludes lowest values of these parameters.
Collecting all our results we finally conclude~:\\
$$ \hbox{a logarithmic rise in }\quad
\ln ^\gamma s ,\quad \gamma\in [0.5,2.20] $$
$$ \hbox{ and a power-like rise in }\quad
s^{\epsilon} ,\quad \epsilon\in [0.01,0.10]$$
are equally probable. By this, we mean that at the same time
the corresponding $\chi^2_{d.o.f}$ are in the range $[1.10 - 1.20]$ and the
corresponding plots of $t=0$ observables give fits which cannot be preferred
"by eye".
As a consequence, we emphasize that it is impossible to settle any definite
preferable values of  $\gamma$ and $\epsilon$, inside the above wide ranges.
This is an illustration of an important feature of the asymptotic total
cross-section, as it may be predicted from actual data above 5 GeV.

As a by-product, it is interesting to note that the dipole Pomeron model
($\equiv $ logarithmic Pomeron with
$\gamma =1$) and the supercritical Pomeron model with very small $\epsilon $ are
similar in describing the available experimental data. Indeed,
one can approximate, for limited values of $s$ and  small $\epsilon $,
the series expansion of the supercritical Pomeron
contribution (8) as
$$P(s) = i \left[g_{0} +
g_{1}\sum_{n=0}^{\infty}\frac{[\epsilon \ln(-is/s_0)]^n}{n!}\right]
\simeq i \left[g_{0} + g_{1}\big[1+ \epsilon \ln(-is/s_0)\big]\right]\eqno(17)$$
and, redefining the Pomeron coupling constants,
identify with the logarithmic Pomeron (7) with $\gamma =1$ ($f$-,
$\omega$-couplings and intercepts for the two descriptions  are unchanged).
Consequently, for any very small $\epsilon$ ($\epsilon <0.01$ for the present
energies)
$$\chi^2 (super crit.\ with\ \epsilon\rightarrow 0)\
\longrightarrow  \ \chi^2 (log.\ with\ \gamma =1) .$$
These properties are still valid in the SE-, EE- and SS-parametrizations and
are checked from the fits and in Figs.1-2. Therefore, one can extend the
interval of allowed $\epsilon$ down to any small non-zero value.

We confirm two main results of Kang and Kim \cite{KK} concerning
the $\gamma$ determination and the failure of the original DL model.
The first one supports a
possible asymptotic cross-section either in $\ln s$ or $\ln^2 s$. The second
one (based on a worse $\chi^2_{d.o.f}$ obtained for DL parametrization) simply
alleviates the need for a modification of the original model but
does not rule out a $s^{\epsilon}$-behavior.
To that respect, we recall that the minor modification we introduced in the
DL parametrization ({\it i.e.} non identical
Reggeon intercepts $\alpha_f(0)\ne\alpha_{\omega}(0)$ and a Pomeron
with a constant term $g_0\ne0$) appears to be one solution for
improving the fit. \\
Such a philosophy has been previously adopted by
Cudell {\it et al.} \cite{CKK}, who also modified the original DL model and
concluded on a fit competitive with any other good fit. \\
The Pomeron intercept we proposed above slighly differs from the central value
of recent fits in \cite{CKK}~:
$\alpha_P(0)= 1+\epsilon \in [1.07,1.11]$. This is not surprising because the
authors uses a parametrization closer than
ours to the original DL model, where [3] $\alpha_P(0)= 1.0808$.
The lower energy cut in the data set we choose
(namely 5 GeV instead of 10 GeV in \cite{CKK}) which is, once again, a
motivation for our
generalization of the DL parametrization  is responsible for that situation.
The Pomeron intercept
is probably underestimated in our fits including only $pp$ and $\bar pp$ data
concentrated at low energies where the Pomeron contribution is expected to be
relatively small. To that respect, one must note following Covolan {\it et al.}
\cite{Cov} that, including in addition $\pi^\pm p$ and $K^\pm p$ data supports
a higher intercept at the Born level 1.104, which is still shifted after
eikonalization up to 1.122.

As an additional remark, we note that increasing the number of parameters is
not necessary to get a better fit, and therefore
there is little place for introducing
an Odderon at t=0~: we do not mean that it
does not exist but simply that the $t=0$--data are not constraining enough to
determine its parameters (similarly, they do not allow to
distinguish $a-$ from $f-$meson and $\rho-$ from $\omega-$meson).

As an example of the analytical fits we obtain, we show
$\sigma_{tot}$ versus $\sqrt{s}$ in Fig.3 and $\rho$ versus $\sqrt{s}$ in Fig.4
for $\gamma=1$ and $\gamma=2$ in (7)
and for $\epsilon=0.07$ in (8). The parameters for these particular cases are
listed in Table 1.
Despite the fact that the weight of the highest energies data are small in the
fit, one obtain a very good agreement at all energies.
In the energy range experimentally known the three sets of results are very
similar~:
the $\gamma=2$ logarithmic Pomeron and the $\epsilon =0.07$
power-like Pomeron give undiscernable curves up to the Tevatron energy,
a visible splitting with the $\gamma=1$ logarithmic Pomeron
occurs only above 500 GeV for the
cross-sections and above 200 GeV for the $\rho$-ratios.
Predictions of these models for the highest energies
are presented in Table 2 and in Fig.5. The three cases exhibited
are in agreement with cosmic ray data \cite{Honda},
but the LHC results would allow a selection.

To summarize our main conclusion~: it seems impossible, at present, to
establish whether a $\ln s$ or a $\ln^2 s$ or an $s^{\epsilon}$ behavior
for the total cross sections is favored by the data.
All the above rises are equally admissible unless a combined fit including also
angular distributions is performed. It is our belief that
different authors will agree with each other only when new
very high energy experimental results will become available and, hopefully,
asymptopia will have been reached.
\bigskip

{\it Acknowledgements} We thank E. Predazzi for his interest and a critical
reading of the manuscript. We are indebted to M. Haguenauer for valuable critics
and suggestions.
One of us (E.M) thanks the Institut de Physique
Nucl\'eaire de Lyon (France) for hospitality during part of this work.
Financial support from the IN2P3 to E.M is gratefully acknowledged.

\newpage

\def\init{\tabskip 0pt\offinterlineskip}

\def\crr{\cr\noalign{\hrule}}

$$\vbox{\init\halign to 12.cm{
\strut#&\vrule#\tabskip=1em plus 2em&
\hfil $#$\hfil&
\vrule#&
\hfil $#$\hfil&
\vrule #&
\hfil $#$\hfil&
\vrule #&
\hfil $#$\hfil&
\vrule #\tabskip 0pt\crr
&&  &&\hbox{{\it logarithmic }}&&\hbox{{\it logarithmic }}&&
\hbox{{\it super.}}&\cr
&&  &&(\gamma=2.0)   && (\gamma=1.0) &&(\epsilon=0.07) &\crr
&&g_f\ (\hbox{ mb})  && 2.385   &&4.559  &&2.971 &\cr
&&\alpha_f(0)\  (\hbox{ GeV}^{-2}) && 0.674  && 0.804  && 0.711 &\cr
&&g_\omega\ (\hbox{ mb})  &&  1.851  &&1.831  && 1.854  &\cr
&&\alpha_\omega(0)\  (\hbox{ GeV}^{-2})&&   0.436  && 0.441  &&0.435      &\cr
&& g_0 \ (  \hbox{ mb})&&  0.9978 &&-1.396  &&-1.009  &\cr
&&g_1     \ (  \hbox{ mb}) &&   0.9114  &&0.2762   &&1.412  &\crr
&&\chi^2_{dof}   && 1.150 &&1.181  && 1.174  &\crr
}}$$

{\large {\bf Table 1 }}

\smallskip
\noindent
Parameters of the analytic amplitude for the $f$-Reggeon (5),
the $\omega$-Reggeon (6) and the Pomeron either logarithmic (7) with fixed
$\gamma=2$, or
$\gamma=1$, or supercritical (8) with fixed $\epsilon=0.07$.
The 3 sets are fitted to the 187 data with $\sqrt{s}\ge 5$ GeV.

$$\vbox{\init\halign to 12.cm{
\strut#&\vrule#\tabskip=1em plus 2em&
\hfil $#$\hfil&
\vrule#&
\hfil $#$\hfil&
\vrule#&
\hfil $#$\hfil&
\vrule#&
\hfil $#$\hfil&
\vrule#\tabskip 0pt\crr
&& &&\hbox{{\it logarithmic }}&&\hbox{{\it logarithmic }}&&
\hbox{{\it super.}}&\cr
&& &&(\gamma=2.0)   && (\gamma=1.0) &&(\epsilon=0.07) &\crr
&&\rho\ (1.8 \hbox{ TeV} ) && 0.138  && 0.121 && 0.140(0.143) &\cr
&&\sigma\ (1.8 \hbox{ TeV } ) \hbox{ in  mb} &&  76.5  &&74.7  && 76.3 &\cr
&&\rho\ (14  \hbox{ TeV} ) && 0.126  && 0.101  && 0.135  &\cr
&&\sigma\ (14 \hbox{ TeV } ) \hbox{ in  mb} &&  108. &&100. && 109. &\crr
}}$$

{\large {\bf Table2}}.

\smallskip
\noindent
High energy $pp$ predictions for the total cross-section $\sigma$ and
$\rho$-ratio,
calculated with the analytic amplitudes and parameters from Table 1.
The $\bar p p$ results are the same, except the $\rho$-value in parentheses.

\newpage

\vskip 2.cm

\begin{center}
\large { \bf Figure captions}
\end{center}

\vskip 1.cm
\noindent
{\bf Fig.1} \\
-a)  $\gamma$-dependence of $\chi^2_{d.o.f}$  for
the analytic amplitude fit (4,7), including
 a logarithmic Pomeron  leading to a total
cross-section rising asymptotically as $\ln^\gamma{s}$. \\
-b)   $\epsilon$-dependence of $\chi^2_{d.o.f}$ for
the analytic amplitude fit (4,8), including
a supercritical Pomeron
leading to a total cross-section rising asymptotically as
$s^\epsilon$.\\
The solid lines correspond to fits performed with the non-filtered data set
between 5 GeV and 1800 GeV;  in dashed lines are the results from similar fits
but omitting data above 546 GeV.

\medskip
\noindent
{\bf Fig.2} \\
-a)  $\gamma$-dependence of $\chi^2_{d.o.f}$  for
 the dispersion relation fit and the logarithmic Pomeron
(11-14,16).\\
-b)   $\epsilon$-dependence of $\chi^2_{d.o.f}$ for
 the dispersion relation fit and the supercritical
Pomeron (11-13,15-16)\\
All the fits are performed with the non-filtered data set.
The solid lines correspond to fits performed within the
ES-parametrization, the dashed lines to SS-parametrization, the dotted lines to
EE-parametrization.

\medskip
\noindent
{\bf Fig.3} \\
Behavior of the cross-sections $\sigma_{pp}(s)$ and $\sigma_{\bar p p}(s)$
as given
from the analytical fits for $\gamma=2$ (solid line), for $\gamma=1$ (dashed
line) and for
$\epsilon=0.07$ (dotted line). The other 6 free parameters are listed in
Table 1.

\medskip
\noindent
{\bf Fig.4} \\
Same as in Fig.3 for the ratios $\rho_{p p}(s)$ and $\rho_{\bar p p}(s)$.

\medskip
\noindent
{\bf Fig.5} \\
 Extrapolated  $\sigma(s)$ and $\rho(s)$ at high energies for the
three analytical fits of Fig.3 ($pp$ and $\bar pp$ are undiscernable).
The cosmic ray data are from \cite{Honda}.

\pagestyle{empty}

\input epsf

\epsfxsize=17 true cm

\centerline{\epsfbox{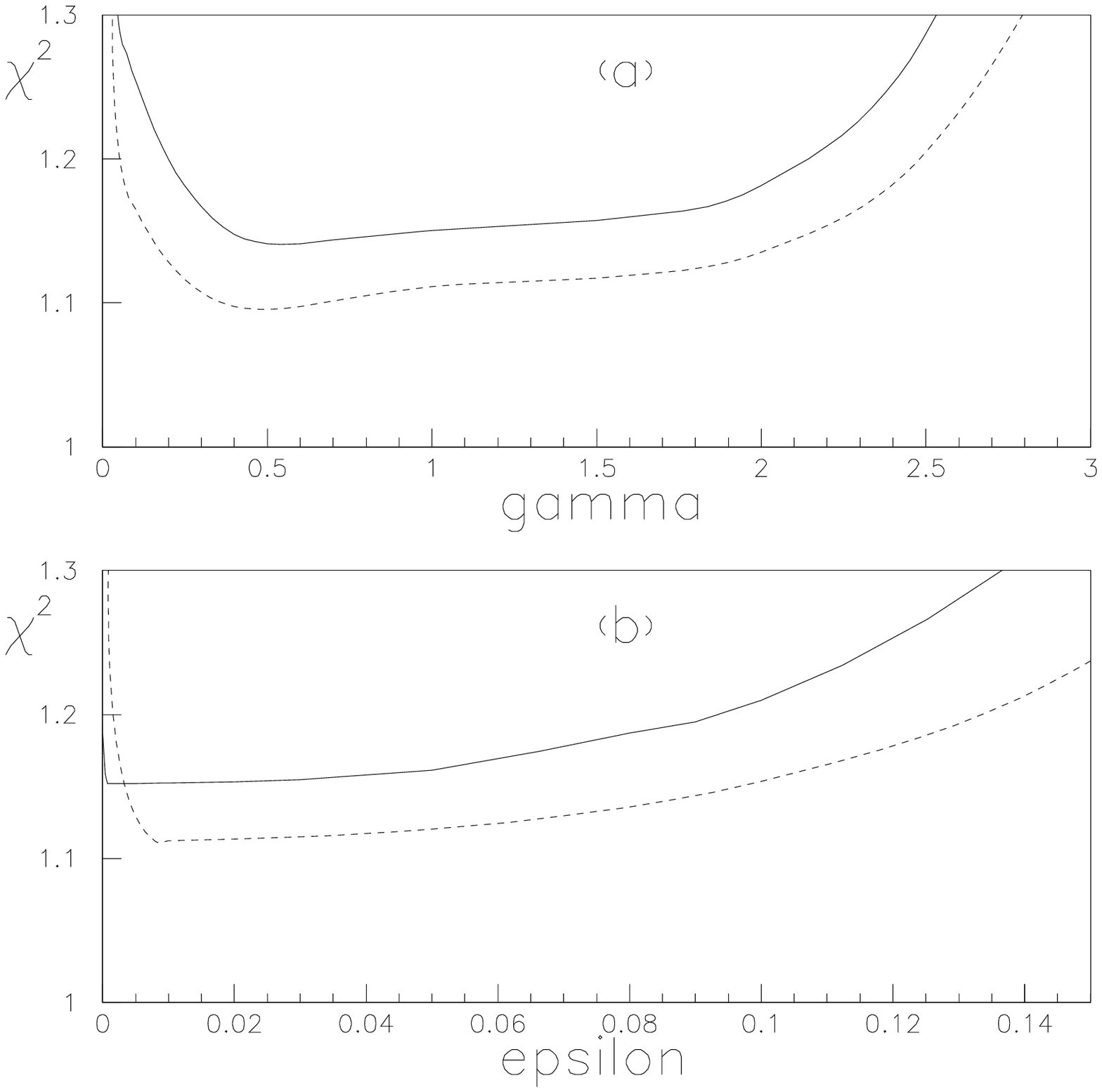}}

\centerline{\epsfbox{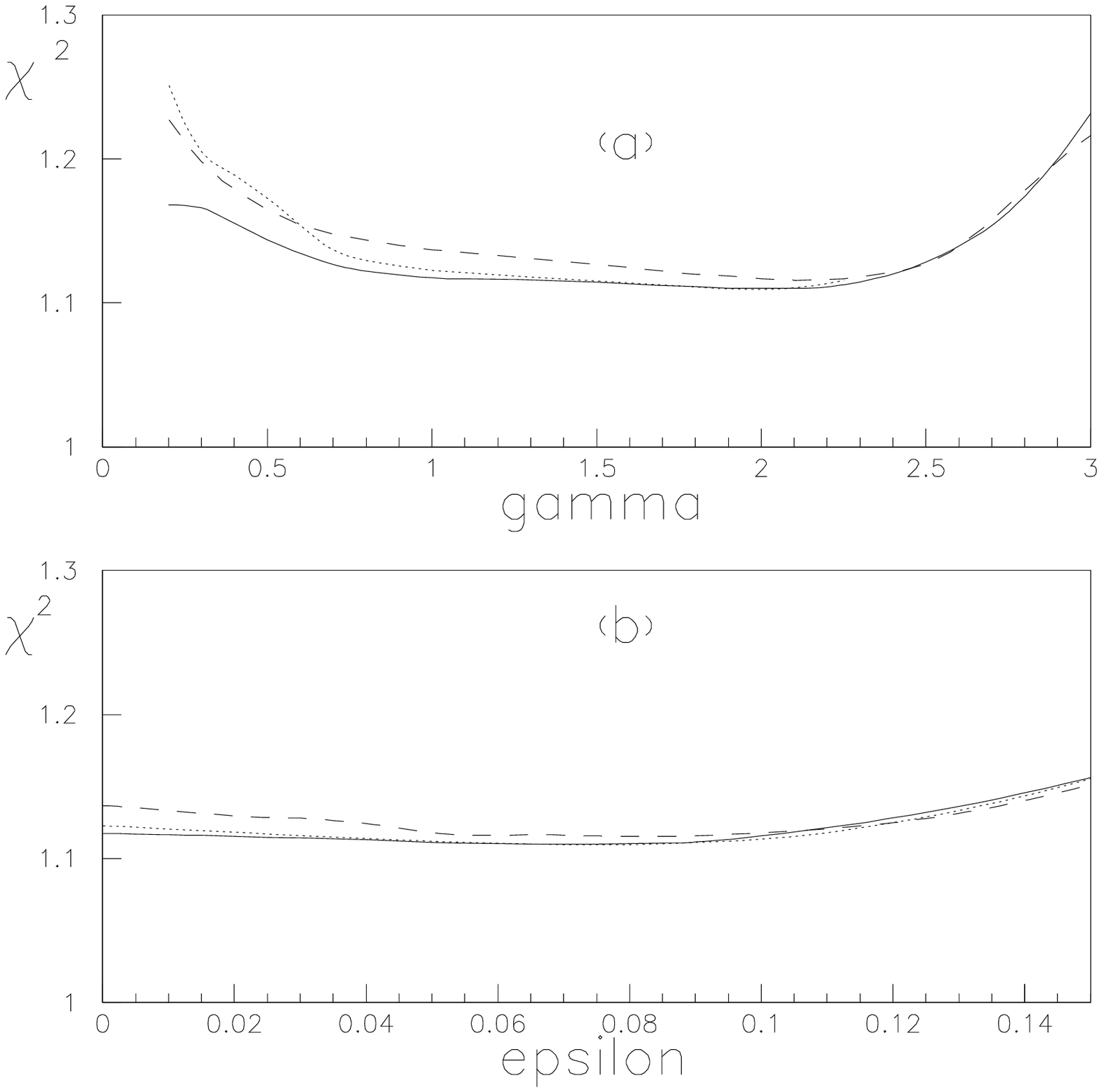}}

\centerline{\epsfbox{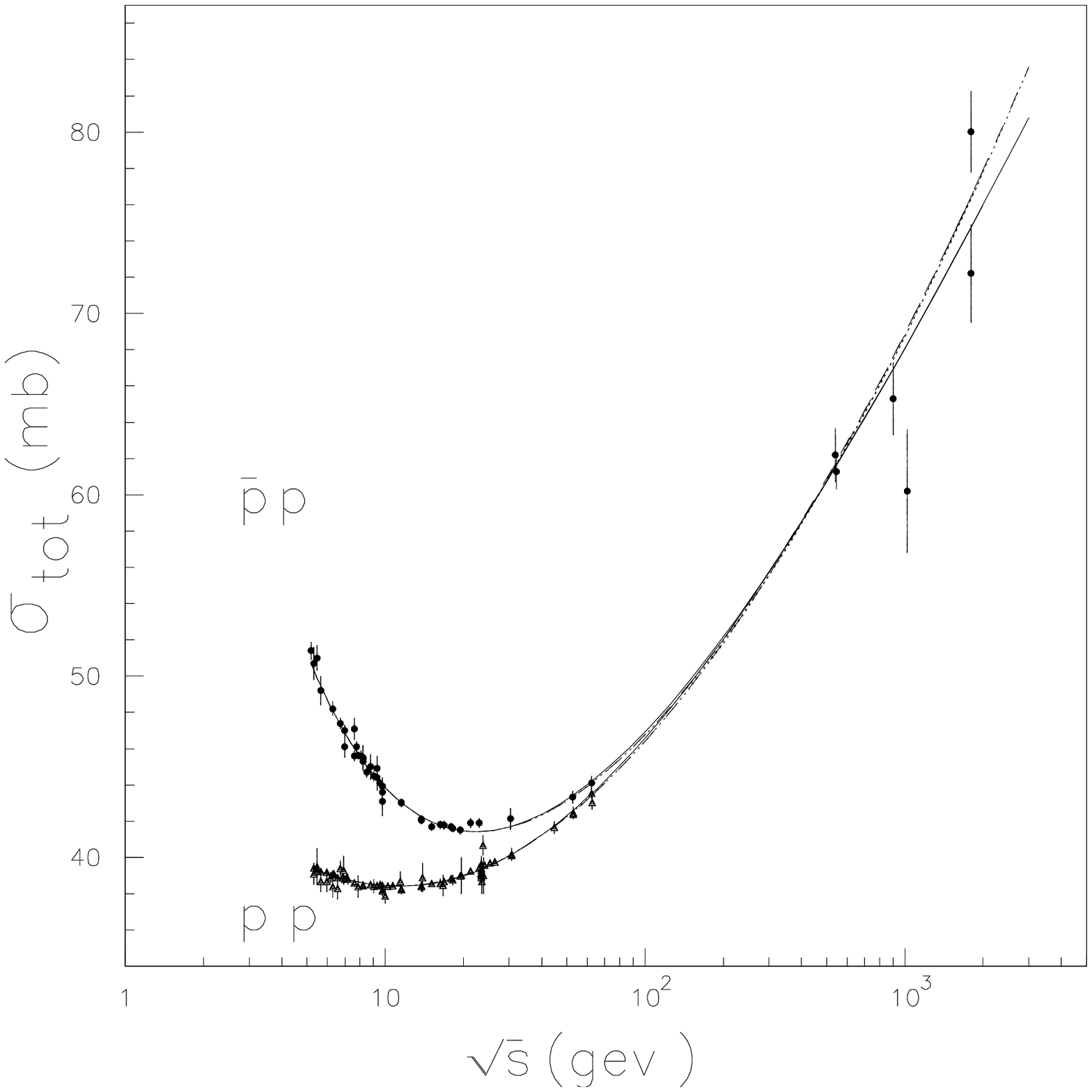}}

\centerline{\epsfbox{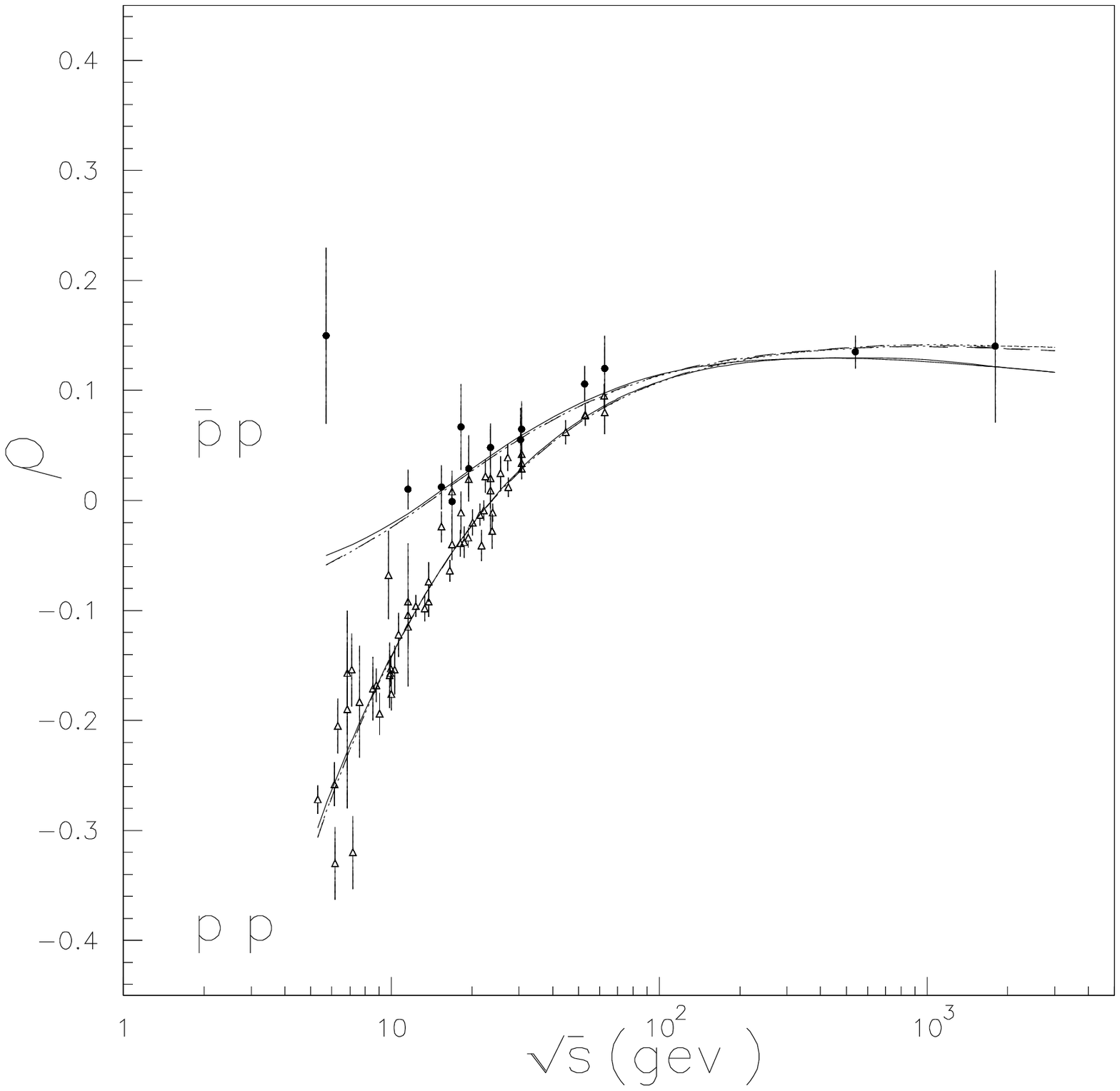}}

\centerline{\epsfbox{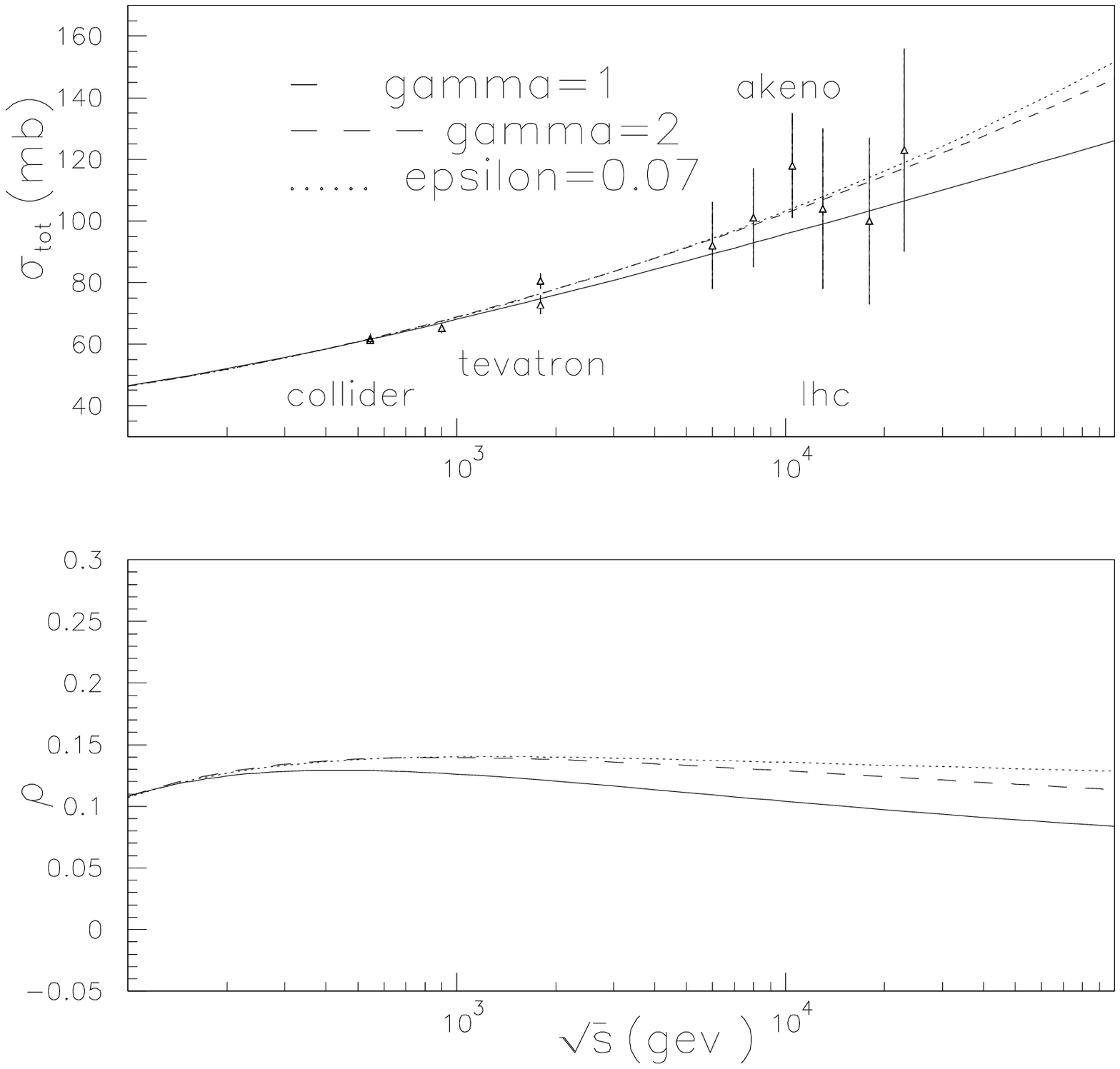}}

\end{document}